# Sliding Flexoelectricity in Two-Dimensional van der Waals Systems

Ri He,[1, 5] Hua Wang,[2] Fenglin Deng,[3] Yuxiang Gao,[3] Binwen Zhang,[4] Yubai Shi,[1, 5] Run-Wei Li,[1, 5, 6] and Zhicheng Zhong[3, 7]

[1]*Key Laboratory of Magnetic Materials Devices & Zhejiang Province Key Laboratory of Magnetic Materials and Application Technology, Ningbo Institute of Materials Technology and Engineering, Chinese Academy of Sciences, Ningbo 315201, China*

[2]*Center for Quantum Matter, School of Physics, Zhejiang University, Hangzhou 310058, China*

[3]*Suzhou Institute for Advanced Research, University of Science and Technology of China, Suzhou 215123, China*

[4]*Fujian Key Laboratory of Functional Marine Sensing Materials, Center for Advanced Marine Materials and Smart Sensors, College of Material and Chemical Engineering, Minjiang University, Fuzhou 350108, China*

[5]*College of Materials Science and Opto-Electronic Technology, University of Chinese Academy of Sciences, Beijing 100049, China*

[6]*School of Future Technology, University of Chinese Academy of Sciences, Beijing 100049, China*

[7]*School of Artificial Intelligence and Data Science, University of Science and Technology of China, Suzhou 215123, China*

**ABSTRACT**

Two-dimensional sliding ferroelectrics, with their unique stacking degrees of freedom, offer a different approach to manipulate polarization by interlayer sliding. Bending sliding ferroelectrics inevitably leads to interlayer sliding motion, thus altering stacking orders and polarization properties. Here, by using machine-learning force field, we investigate the effects of bending deformation on geometries, stackings, energies, and polarizations in sliding ferroelectric bilayer h-BN and 3R-MoS$_2$. We predict that bent ferroelectric bilayer forms irreversible kinks instead of arc when the bending angle exceeds a critical value. We demonstrate that the kinks originate from the competition between bending energy and interlayer van der Waals energy. The kink contains a ferroelectric domain wall that reverses the polarization, effectively inducing a flexoelectric effect. We term this phenomenon "sliding flexoelectricity" to distinguish it from conventional strain-gradient-induced flexoelectricity.

heri@nimte.ac.cn

Sliding ferroelectricity has been recently observed in artificially stacked two-dimensional (2D) WTe$_2$[1], h-BN[2,3], 3R-MoS$_2$[4], and MoTe$_2$[5], where the out-of-plane polarization stems from the interlayer charge transfer due to their stacking degrees of freedom. Their polarization can be switched through the interlayer sliding with an ultralow barrier[6,7]. Especially, recent theoretical and experimental researches both show that sliding ferroelectricity exhibits ultrafast switching dynamics and excellent anti-fatigue performance because of the nature of the sliding ferroelectric domain wall[4,8,9], making sliding ferroelectrics promising candidates for next-generation functional devices, such as nonvolatile memories, sensors, and wearable electronics.

Flexoelectricity, defined as the coupling between bending deformation and polarization property, plays a significant role in ferroelectrics[10], where it can enhance or reverse the polarization and even manipulate ferroelectric domain structures[11-14]. Like most vdW materials, sliding ferroelectrics have high fracture stress and low bending stiffness, enabling them easy to achieve significant out-of-plane bending, corresponding to strain gradients 2 to 3 orders of magnitude higher than those achieved via heteroepitaxy in oxides thin films[15]. Thus, flexoelectricity is expected to be extremely strong in sliding ferroelectrics. More importantly, sliding ferroelectrics, due to their low interlayer sliding barriers and high in-plane stiffness[6,7], inevitably experience interlayer sliding when bent, thereby altering their stacking order and polarization properties. Consequently, the multinomial coupling involving bending, interlayer sliding, stacking order, and polarization must be considered in the bending of sliding ferroelectrics. It implies that the conventional gradient-strain-induced flexoelectricity theory may be inapplicable to sliding ferroelectrics. However, the interlayer sliding and stacking degrees of freedom are often neglected in current density functional theory (DFT) calculations and Landau-Ginzburg-Devonshire model simulations of flexoelectricity in 2D ferroelectrics[16-18], leaving a fundamental question about the nature of the flexoelectric effect in sliding ferroelectrics unanswered.

To answer the above question with quantum-mechanical accuracy, it has become imperative to employ atomic simulations with first principles DFT calculation.

However, studying complex strain states and bending deformations, which requires breaking translational symmetry, poses a significant challenge for DFT[13]. Recent advances in machine learning interatomic potentials model open avenues for investigating the flexoelectric effect in 2D vdW materials[16,19]. In recent works, we have developed the machine learning deep potential (DP) models for sliding ferroelectric bilayer h-BN and 3R-MoS$_2$ using a training dataset obtained from the DFT calculations[4,8]. Our DP models have been demonstrated excellent agreement with DFT results across a wide range of properties.

Herein, we further extend our training datasets of bilayer h-BN and 3R-MoS$_2$ to include the additional bent configurations. The methodology details and further validation of machine learning potentials accuracy can be found in the Supplemental Material [20]. Using the DP models, we investigate the effect of bending deformation on geometries, stackings, energies, and polarizations in sliding ferroelectric bilayer h-BN and 3R-MoS$_2$. We find that the bilayer sliding ferroelectric system produces the irreversible kinks when the bending angle exceeds 12˚. Our analytical model demonstrates that the kinks originate from the competition between bending energy and interlayer vdW energy. The kinks contain ferroelectric domain walls that reverse the out-of-plane polarization. As a result, under bending deformation, interlayer sliding can induce various intermediate stacking configurations and manipulate ferroelectric domain structures, a phenomenon we term "sliding flexoelectricity". Our work provides an atomic-scale understanding of flexoelectric effect in sliding ferroelectrics, and the proposed sliding flexoelectricity can be extended to other multilayered 2D vdW materials.

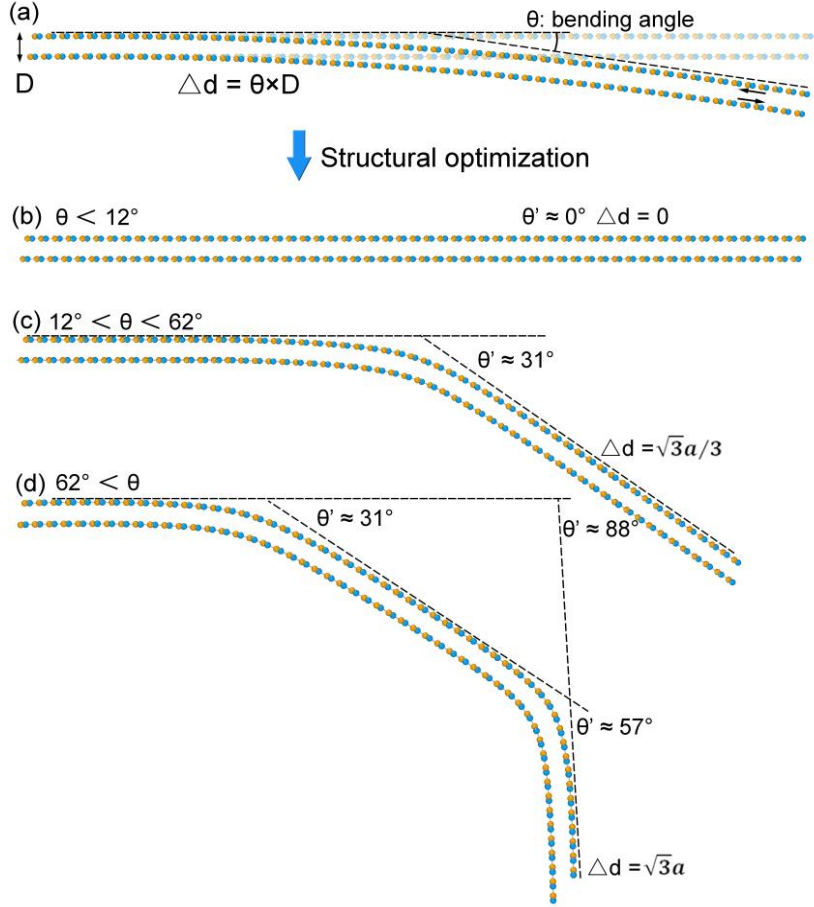

Figure 1. Atomic structure of AB domain bilayers h-BN with different bending angles. (a) Flat sliding ferroelectricity film was artificially bend along the armchair direction with specific bending angle θ. The interlayer distance and in-plane lattice in each layer remain constant during bending deformation. The bending induced sliding distance Δd can be written as a function of bending angle θ: Δd = θ × D, where D is interlayer distance. (b-d) Optimized structures of bilayer with different initial bending angles obtained from deep potential model.

To begin, we examine the stacking modes of bilayer h-BN. These stacking modes can be divided into parallel and antiparallel twist orientations. Only the parallel configuration exhibit sliding ferroelectricity because the lack of inversion symmetry allows for interlayer charge transfer, resulting in a spontaneous out-of-plane polarization[7]. The parallel configurations have energetically favorable polar-AB, -BA stacking orders, and unstable AA, saddle point (SP) stacking orders. We start from the stable AB single-domain h-BN bilayer sheet. To induce bending deformation, we artificially bend the flat sheet along $x$ axis with specific angle θ as illustrated in Fig. S1[20]. The characteristic length of bent region is 12.3 nm, and the residual regions remain flat. After bending deformation, the bilayer sheet forms an arc, as shown in Fig. 1a. Due to the high stiffness of the in-plane hexagonal structure of h-BN, we regard the interlayer distance and in-plane lattice as constants under bending. As a result, bent region suffers from continuously interlayer sliding, and sliding distance Δd of terminal side can be written as a function of bending angle θ: Δd = θ × D, where D is the interlayer distance of bilayer. It should be noted that the Δd is independent of the characteristic length of the bent (arc) region. The atomic structure of bilayer with a bending angle θ is shown in Fig. S1b[20].

    The artificially bent bilayer structures are energetically unstable. The DP model for the atomic relaxation leads to the optimized bilayer structures with different bending angles. We define the bending angle of optimized structures as θ'. After atomic relaxation, as shown in Fig. 1b-d, we observe arc structure transform to three classes of optimized structures: flat, θ' = 31° kink, and two kinks of θ' = 31° and 57°. These structures depend on their bending angle θ. We investigate the effect of bending angle θ on the morphology of θ'. In Figure 2b, we plot the optimized θ' as a function of bending angle θ (0° < θ < 100°). Three discrete stages of θ' are visible in the figure: (i) lower bending angles θ < 12° produce a flat plane (θ' = 0), (ii) bending angles 12° < θ < 62° lead to a kink of θ' = 31°, and (iii) larger bending angles θ > 62° lead to two kinks of θ' = 31° and 57° in close proximity. This interesting phenomenon is different from the elasticity observed in conventional elastic 2D material sheets, where the bent morphology recovers to flat immediately after bending deformation is removed. The origin of this critical bending angles θ of 12° and 62° can be understood by examining the stacking energy. Fig. 2a plots the stacking energy landscape of interlayer sliding pathway during bending deformation. It shows that the AB and BA stackings have local minimum energy, while the SP and AA stackings have local maximum energy. It implies that the SP and AA stackings are unstable and tend to decay into more energetically favorable AB or BA stackings. As mentioned above, the sliding distance Δd shows a linear dependence on the bending angle θ. The interlayer distance of bilayer h-BN is 3.25 Å. The formations of SP and AA stacking correspond to Δd = 0.71 Å (θ = 12°) and

2.84 Å ($\theta = 62°$). As shown in Fig.2, during the bending AB stacking domain sheet, if the bending angle $\theta$ is less than 12°, the stacking order does not exceed the SP configuration, and the optimized structure will reverts to AB stacking (see the green balls in Fig. 2a), resulting in the flat sheet ($\theta' = 0$). If the bending angle is 12° < $\theta$ < 62°, the structure will transform to BA stacking ($\Delta d = \sqrt{3}\boldsymbol{a}/3$), associated with a domain wall separating AB and BA stacking domains. While the bending angle is large than 62°, the optimized structure will recover to AB stacking ($\Delta d = \sqrt{3}\boldsymbol{a}$). As a result, regardless of the bending angle $\theta$, after relaxation, the sliding ferroelectric h-BN can form the three discrete states. We note that a bending induced kink phenomenon was also recently observed in multilayer α-In$_2$Se$_3$[21]. Such kinks are purely driven by bending energy, which is entirely different with the nature of sliding mechanism in this work.

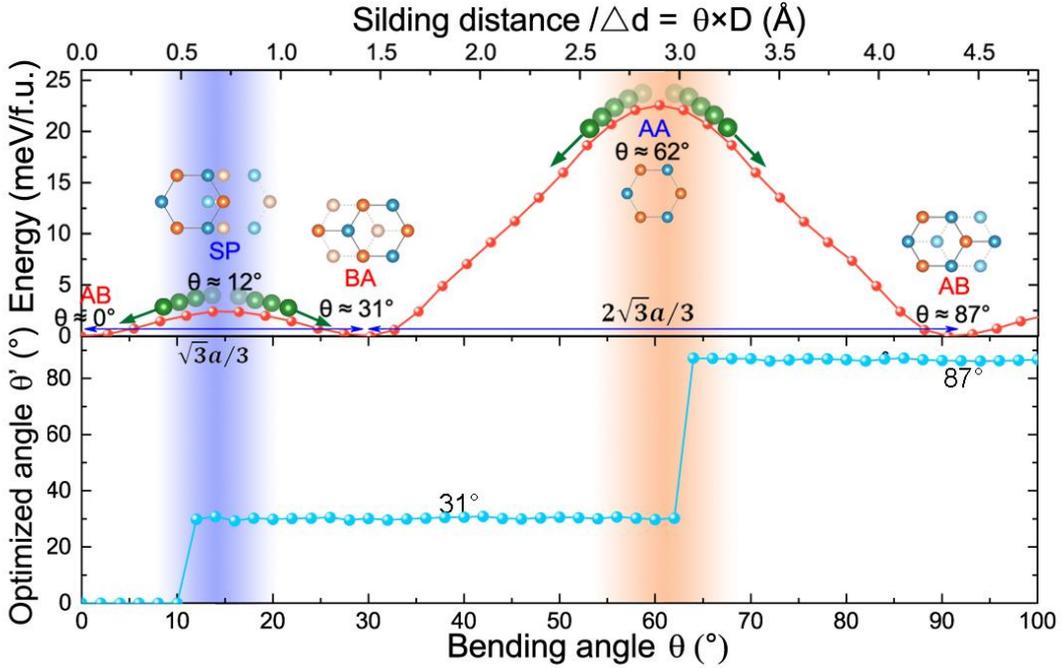

Figure 2. Deep potential predicted energies and structures for kinks formation. (a) Stacking energy landscape as a function of interlayer sliding pathway induced by bending deformation. The AB and BA stackings have local minimum energy, while the saddle point (SP) and AA stackings have local maximum energy. The falling green balls indicate that the SP and AA stackings are quite unstable and tend to decay into more energetically favorable AB or BA stackings. $\boldsymbol{a}$ is the in-plane lattice constant of bilayer unit cell. (b) Plot of optimized structures angle $\theta'$ versus bending angle $\theta$.

Next, we focus on the ferroelectric domain structure of the bent bilayer h-BN.

Without deformation, the equilibrium structure is a single AB domain with a uniform out-of-plane polarization. We plot the polarization switching in the sliding ferroelectric before and after bending deformation. Figure 3a shows a 31° kink (corresponding to figure 1c) in the bilayer forms a Néel-type domain wall, where the polarization rotates smoothly with nearly constant from upward to downward by sliding $\sqrt{3}a/3$, producing a new ferroelectric domain (for details calculations, see Supplemental Material[20]). The domain wall of 31° kink has a SP stacking configuration, which creates pure in-plane electric dipoles. This result is consistent with our recent calculations of 90° sliding ferroelectric domain wall[8]. While another 57° kink induced by a larger bending forms an Ising-like domain wall, collinear with a vanishing polarization at the wall center, as shown in Fig. 3b. The domain wall of 57° kink has AA stacking, which rotates polarization by sliding motion of $2\sqrt{3}a/3$ (Fig. 3c). The inversion symmetric AA stacking is non-polar. In addition, these kinks contain topological line defects, which originates from the difference in atomic number between upper and lower layers induced by interlayer sliding. Consequently, these kinks possess extremely high strain gradients.

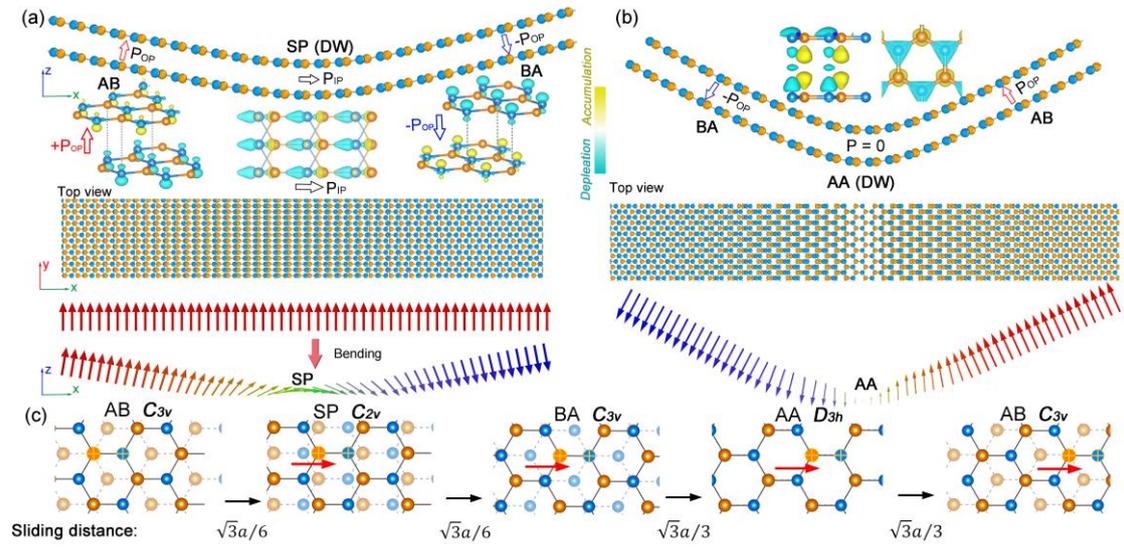

Figure 3. Side and top views of atomic configuration and polarization texture in bent bilayer h-BN. (a) 31° and (d) 57° kinks form a Néel-type and Ising-type ferroelectric domain walls, which reverse the polarization. (c) A relative interlayer sliding by bending deformation results in a cyclic switching between AB, SP, BA, and AA stacking configurations. Red arrows represent the sliding direction.

The kinks induced domain wall reverse or eliminate the polarizations, making an effective flexoelectric effect. Interestingly, such flexoelectric effect in sliding ferroelectrics is strikingly different from the conventional flexoelectric effects that previously reported in intrinsic 2D ferroelectrics, such as $In_2Se_3$ and $CuInP_2S_6$, where the polarization always points outward with respect to the curve[16-18]. For instance, when there is an upward bending in $CuInP_2S_6$ or $In_2Se_3$ flakes, the upper part of the film is elongated, while the lower part is compressed, and such strain gradient propels vertical displacements of Cu or In ions into position of upward polarization. When it is bent downward, the opposite occurs. In contrast, our calculation results show that bending induce an interlayer sliding without any vertical ion displacements in bilayer. The interlayer bending induced kinks that reverse or eliminate the polarization is the dominant mechanism for the flexoelectric effects in sliding ferroelectrics, a phenomenon we term "sliding flexoelectricity". To verify the universality of sliding flexoelectricity, the similar kinks and domain structures are found in bent sliding ferroelectric bilayer 3R-$MoS_2$ as illustrated in Fig. S10[20]. The difference is that the angles of optimized kinks are 17.5° and 34°, almost half of those in h-BN. This is because the bending angle is inversely proportional to the interlayer distance ($\theta = \Delta d/D$), and the interlayer distance of 3R-$MoS_2$ (6.14 Å) is almost twice the 3.25 Å of h-BN (see Table S1[20]).

Rippling deformation is also a common method to investigate flexoelectricity in atomic simulations, thus, a rippling deformation was applied in the bilayer to create a sinusoidal-shaped model with periodic boundary conditions. After fully atomic relaxation by DP model, the geometry changes from arc to kink (Fig. S2[20]). Meanwhile, a sliding induced phase transition from the AB to the BA stacking occurs in the middle of the bilayer. The kinks are very stable against thermal fluctuations even at high temperature of 1100 K (Fig. S11[20]). This robustness of the kinks and sliding ferroelectric domain confirms a recent theoretical prediction of a high Curie temperature[22]. These observations are in agreement with the artificially bent mode results in Fig. 1. Therefore, understanding why a bent sliding ferroelectric bilayer produces atomically sharp kinks instead of arc is an interesting question.

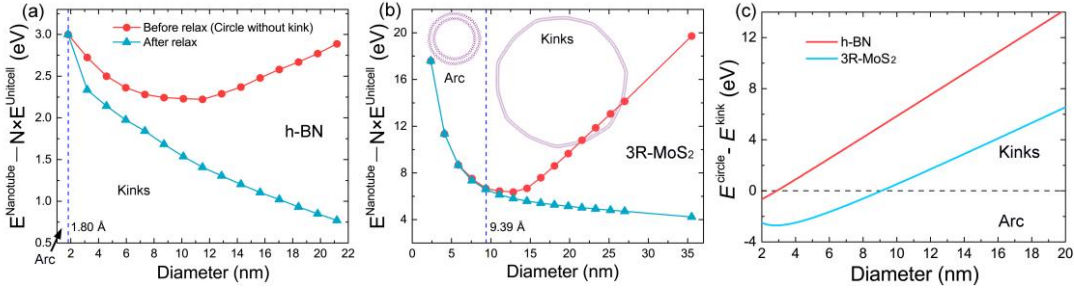

Figure 4. Nanotube formation energies of (a) h-BN and (b) 3R-MoS$_2$ as a function of diameter before and after atomic relaxation by DP model. The formation energies are defined as the energy difference between nanotube and AB stacking bilayer with same number of atoms. (c) The energy difference between circular nanotube ($E^{\text{circle}}$) and kinked nanotube ($-E^{\text{kink}}$) with as a function of diameter calculated by analytical model of Eq. 1.

To reveal the energetic origin of kinks in bent sliding ferroelectrics, we investigate the effect of curvature on kink formation in nanotube. Nanotube is a common strategy to explore the bending effect in 2D materials, which can easily control the curvature and bending energy by variation of diameter[23]. Here, we consider simple armchair double-walled nanotubes of h-BN and 3R-MoS$_2$. In h-BN nanotubes, the difference between inner and outer tubes is eight formula units, because the corresponding interwall distance close to the bilayer interlayer distance. The difference in unitcells between inner and outer tubes leads to continuous variation of stacking orders. We find that when the diameter of h-BN (3R-MoS$_2$) nanotube is larger than 1.8 nm (9.4 nm), it produces several kinks instead of a perfect circle as shown in Fig. S3 and S4[20], which is identical to the recent theoretical work[24]. It indicates that the kinks formation of atomic reconstruction can lower the energy of nanotube. Here, we defined the nanotube formation energy as a function of diameter ($d$): $E^{\text{Nanotube}}(d) - N \times E^{\text{Unitcell}}(d)$, where $E^{\text{Nanotube}}$ and $E^{\text{Unitcell}}$ represent the DFT total energy of nanotube and AB stacking bilayer unitcell. N is the number of unit cells that contained in the nanotube. We plot the energy of nanotube before and after atomic relaxations. Before full relaxation, the nanotubes maintain perfect circles, and they transform into kinked morphology when the diameter exceeds a critical value. The resulting formation energy-versus-diameter diagram of nanotube h-BN and 3R-MoS$_2$ are shown in Fig. 4a and b. It shows that the kinked nanotube is energetically favored when diameter exceeds a critical value. As diameter is decreased, the optimized structure maintains perfect circle. And the energy of optimized structure is almost the same as the one of initial circular structure, indicating the circle structure is energetically stable. In addition, the results show that the formation energy of circular nanotube firstly decreases and then increases linearly with the diameter increases. The initial decrease of formation energy is due to the reduction of bending energy. Then, the linear increase of formation energy originates from the vdW stacking energy increase linearly with diameter increasing. In contrast, the formation energy of kinks structure initially decreases monotonously, and then tends to converge to a certain value. This value is equal to the sum of formation energies of kinks.

The kinks formation is governed by the competition between interlayer vdW interaction energy ($E_{vdW}$) and bending deformation energy ($E_{bending}$). Therefore, the formation energy ($E$) of double-walled nanotube can be written as:

$$E = E_{vdW} + E_{bending} \quad (1)$$

The formation energy of circular nanotube ($E^{circle}$) and kinked nanotube with kinks ($E^{kink}$) can be written as a function of radius of nanotube:

$$E^{circle} = 2\pi r \int_0^{2\pi} u[\psi(\theta)] \, d\theta + \frac{2\pi D_m}{r} \quad (2)$$

$$E^{kink} = \frac{\pi(u_{SP} + u_{AA})}{K} + 2\pi D_m K \quad (3)$$

where $u$ is the interlayer vdW energy density (energy per formula unit). $\psi(\theta)$ is the local stacking configuration at polar coordinates $\theta$. $u_{SP}$ and $u_{AA}$ are the barriers height at SP and AB stacking configurations. $D_m$ denote the out-of-plane bending modulus of 2D monolayers. $K$ is the curvature of kinks. Defining energy difference $\Delta E = E^{circle} - E^{kink}$ allows a direct comparison of the thermodynamic stability between circular and kinked nanotubes with diameter. A positive value of $\Delta E$ indicates the kinked structure is energetically more favorable than the circular one.

We extract the model parameters in equation set from the DFT calculations (see Table S1[20]). Then we plot the $\Delta E = E^{\text{circle}} - E^{\text{kink}}$ in Figure 4c according to Eqs. 2 and 3. It gives the critical diameters of 2.8 nm (h-BN) and 9.1 nm (3R-MoS$_2$) above which the kinked nanotube is more stable thermodynamically than the circular one. The critical diameters from the energy model of Eq. 1 are consistent well with the predictions of the DP model (1.8 nm for h-BN and 9.39 nm for 3R-MoS$_2$). According to energy model, it can be concluded that kinks can lower the vdW interaction energy, but increase the bending deformation energy. When the diameter of nanotube is extremely large, the effect of vdW interaction energy is significant, while the bending deformation energy can be neglected. Thus, the formation of kinks can lower the energy of bent system.

In summary, using the machine-learning potentials, we explore the effect of bending deformation on geometric structures, stacking orders, energies, and polarizations in sliding ferroelectric. We predict that bending in sliding ferroelectric bilayers produce irreversible 31° and 57° kinks instead of arc. The state of kinks is intricately linked interlayer sliding distance induced by bending deformation. We also show that the kinks contain Néel- or Ising-like ferroelectric domain walls that reverse the ferroelectric domain, which makes an effective flexoelectric effect. Such effect is different from the conventional gradient strain-induced flexoelectricity, therefore we term this phenomenon "sliding flexoelectricity". Finally, we demonstrate that the kinks originate from the competition between bending energy and interlayer van der Waals energy by using an energy model. Our findings suggest an interlayer sliding mediated flexoelectricity in sliding ferroelectrics, which could accelerate the development of slidetronics-based functional devices.

**Acknowledgments**

This work was supported by the National Key R&D Program of China (Grants No. 2021YFA0718900 and No. 2022YFA1403000), the National Nature Science

Foundation of China (Grants No. 11974365, No. 12204496, and No. 12304049), and the Zhejiang Provincial Natural Science Foundation (Grants No. Q23A040003 and No. LDT23F04014F01).

# Supplemental Material

# Sliding Flexoelectricity in Two-Dimensional van der Waals Systems

Ri He,[1, 5] Hua Wang,[2] Fenglin Deng,[3] Yuxiang Gao,[3] Binwen Zhang,[4] Yubai Shi,[1, 5] Run-Wei Li,[1, 5, 6] and Zhicheng Zhong[3, 7]

[1]*Key Laboratory of Magnetic Materials Devices & Zhejiang Province Key Laboratory of Magnetic Materials and Application Technology, Ningbo Institute of Materials Technology and Engineering, Chinese Academy of Sciences, Ningbo 315201, China*

[2]*Center for Quantum Matter, School of Physics, Zhejiang University, Hangzhou 310058, China*

[3]*Suzhou Institute for Advanced Research, University of Science and Technology of China, Suzhou 215123, China*

[4]*Fujian Key Laboratory of Functional Marine Sensing Materials, Center for Advanced Marine Materials and Smart Sensors, College of Material and Chemical Engineering, Minjiang University, Fuzhou 350108, China*

[5]*College of Materials Science and Opto-Electronic Technology, University of Chinese Academy of Sciences, Beijing 100049, China*

[6]*School of Future Technology, University of Chinese Academy of Sciences, Beijing 100049, China*

[7]*School of Artificial Intelligence and Data Science, University of Science and Technology of China, Suzhou 215123, China*

# S1. Methods

Machine-learning potential of bilayer h-BN and 3R-MoS$_2$

Machine-learned potentials provide a high-dimensional fit of the potential energy surface of materials[1]. The basic idea of machine-learning potential is construction of deep neural network and to fit the DFT calculation data of abundant configurations. For a well-trained DP model, given any large-scale configuration, it can figure out the corresponding total energy and atomic forces at DFT-level accuracy. Recently, we developed a machine-learning-based deep potential (DP) models for bilayer h-BN and 3R-MoS$_2$ using training dataset from DFT calculation[2,3]. A large number of representative configurations in training dataset were obtained by performing concurrent learning procedure[4]. In this work, our focus is on the studying the electromechanical property of bilayers at bending deformation conditions, the bending break the crystal periodicity of supercell. Therefore, the configurations of nanoribbon and bent bilayer are added in the training datasets.

The DP model is trained using these configurations and corresponding PBE-based DFT energies, with fitting deep neural network of size (240, 240, 240). The cutoff radius of the model is set to 6.5 Å for neighbor searching, while the smoothing function starts from 2.5 Å. The DEEPMD-KIT code is used for training of DP model[5]. The DP compression scheme was applied in this work for accelerating the computational efficiency of the DPMD simulations[6]. The final training datasets and DP models to reproduce the results in this paper are available in the DP-library website (https://www.aissquare.com).

**DFT calculations**

We use the Vienna ab initio simulation package (VASP)[7] with projector augmented wave method and the generalized gradient approximation (GGA) with Perdew-Burke-Ernzerh of exchange-correlation functional[8]. We adopt a plane-wave cutoff energy of 600 eV in the structural relaxation calculations. A large distance of $c >$ 20 Å along the out-of-plane direction eliminates interlayer interactions. To obtain an accurate deep potential model for layered *h*-BN, we employ the Van der Waals correction with optB86b functional in DFT calculations[8]. The Brillouin zone was sampled with a 6 × 6 × 1 k-point grid for the 3 × 3 × 1 supercell. The convergence criterion of energy was $10^{-6}$ eV. The optimized lattice constant of AB stacking h-BN (3R-MoS$_2$) is $a = b$ = 2.510 Å (3.166 Å) with interlayer distance 3.256 Å (6.14 Å). The polarizations of various stacking orders are calculated by the Berry phase method, and the results are show in Fig. S9. To reproduce the interlayer sliding motion between AB, SP, BA, and AA stacking, the climbing image nudged elastic band (CINEB) approach was used to find the stacking energy landscape[9]. The DFT calculated bending modulus for h-BN and 3R-MoS$_2$ are 3.6 eVÅ$^2$ and 43.3 eVÅ$^2$ (see Table 1), in good agreement with previous works[10,11].

**Molecular dynamics simulations**

The MD simulations were carried out using LAMMPS code with periodic boundary conditions[12]. The atomistic interactions of bilayer systems are described by the developed DP models. The Nose-Hoover thermostat is employed to control temperature[13]. The velocity Verlet algorithm is used for integrating the equations of motion with a time step of 1.0 fs in all MD simulations. The DPMD simulations start with 2 × √3 cell building blocks.

**Construction of atomic model**

To address the non-periodic issues of bending deformation in bilayer sliding ferroelectrics, we construct three different bending models, including nonperiodic nanoribbon, periodic sinusoidal shape, and nanotube models. For nanoribbon model, 2 × √3 cell building blocks was used to was construct the 2D superlattice with 160 × 4 cells as shown in Fig. S1. The vacuum region was used in *x*-[1 0 0] and *z*-[0 0 1] direction and periodicity was used in *y*-[0 1 0] direction. The bending strategy in nanoribbon was illustrated in Fig. S1a. We artificially bend the flat sheet along the armchair direction with different radii circles. The characteristic length of bent region is ~12.3 nm, and the residual regions keep flat. After bending deformation, the bilayer sheet transforms to arc as shown in Fig. S1c. For sinusoidal-shaped model, the displacement of each atom in out-of-plane direction can be described by equation of $\sin(2\pi x)$, where $x$ is the atomic position in the x-axis direction, $m$ is the rippling amplitude (Fig. S2). For nanotube model, double-walled armchair nanotubes with various sizes were used in simulations (Fig. S3 and S4). It is found that when the difference of inner and outer tubes is eight formula units for h-BN, it has the lowest free energy.

## S2. Validation of the Machine-Learning Deep Potential (DP)

The training datasets are generated by a concurrent learning procedure automatically, which cover all relevant configuration space. For details, please refer to our recent works[2,3]. The performance of the DP model is validated through the comparison of the energies and forces from DFT calculations in the testing dataset. The energies and forces predicted by the DP model versus DFT calculated energies and forces are plotted in Fig. S5, the mean absolute error (MAE) of energes $|E^{DFT} - E^{DP}|$ and atomic forces $|f^{DFT} - f^{DP}|$ between DP and DFT are 0.60 meV/atoms and 0.052 eV/Å for h-BN, and 0.75 meV/atoms and 0.02 eV/Å for 3R-$MoS_2$, respectively, suggesting the high accuracy of the DP models. Significantly, the DP models can accurately reproduce the energy barrier under interlayer sliding process as illustrated in Fig. S7, again demonstrating excellent agreement between DP and DFT results. In addition, we relax the nanotube structures with diameter of 1.6 nm and 3.2 nm. The atomic structures for the DP relaxations agree well with DFT results, and the corresponding results are presented in Fig. S8. We also calculate the phonon dispersion relations of sliding ferroelectric h-BN and 3R-$MoS_2$ bilayers by the DP model and DFT in Fig. S6. The calculated phonon dispersions by the DP model is found to be almost overlapping with DFT curves, demonstrating good description of the second-order derivative information around local minima of ferroelectric phase. Moreover, the phonon dispersion demonstrated the dynamical stability of AB stacking bilayer at ambient condition, with no negative phonon frequencies. Overall, the DP model accurately predicted the thermodynamic and kinetic properties of h-BN and 3R-MoS$_2$ sliding ferroelectrics.

## S3. Polarization property of kinks

The in-plane and out-of-plane polarization along the sliding pathway are calculated by Berry phase method based on DFT. We plot the in-plane and out-of-plane polarization profiles along the sliding pathway in Fig. S9. The AB stacking bilayer h-BN with $C_{3v}$ symmetry has a pure out-of-plane polarization of 1.46 pC/m. While the intermediate saddle-point (SP) stacking with $C_{2v}$ symmetry exhibits a pure in-plane polarization of 1.53 pC/m. The another energetically unfavorable AA stacking with $C_{3h}$ symmetry is nonpolar. These polarization values are consistent with our previous study[2,3]. The bending deformation can induce various stacking mode in the kinks. Mapping from polarization profiles in Fig. S9 to stacking mode bent sample, we obtain the polarization textures in Fig. 3.

Table SI. The DFT calculated out-of-plane bending modulus $D_m$, curvature of kink $K$, and height at SP and AB stacking $u_{SP}$ and $u_{AA}$.

|  | h-BN | 3R-MoS$_2$ |
|---|---|---|
| $D_m$ (eVÅ$^2$) | 3.6 | 34.3 |
| $K$ (/Å) | 0.002 | 0.0012 |
| $u_{SP}$ (eV) | 2.395 | 7.65 |
| $u_{AA}$ (eV) | 22.54 | 34.765 |
| Interlayer distances D (Å) | 3.25 | 6.14 |
| Lattice constant (Å) | 2.51 | 3.16 |
| Sliding distances from AB to BA stacking (Å) | 1.45 | 1.827 |
| Bneding angle (°) | 31°/57° | 17.5°/34° |

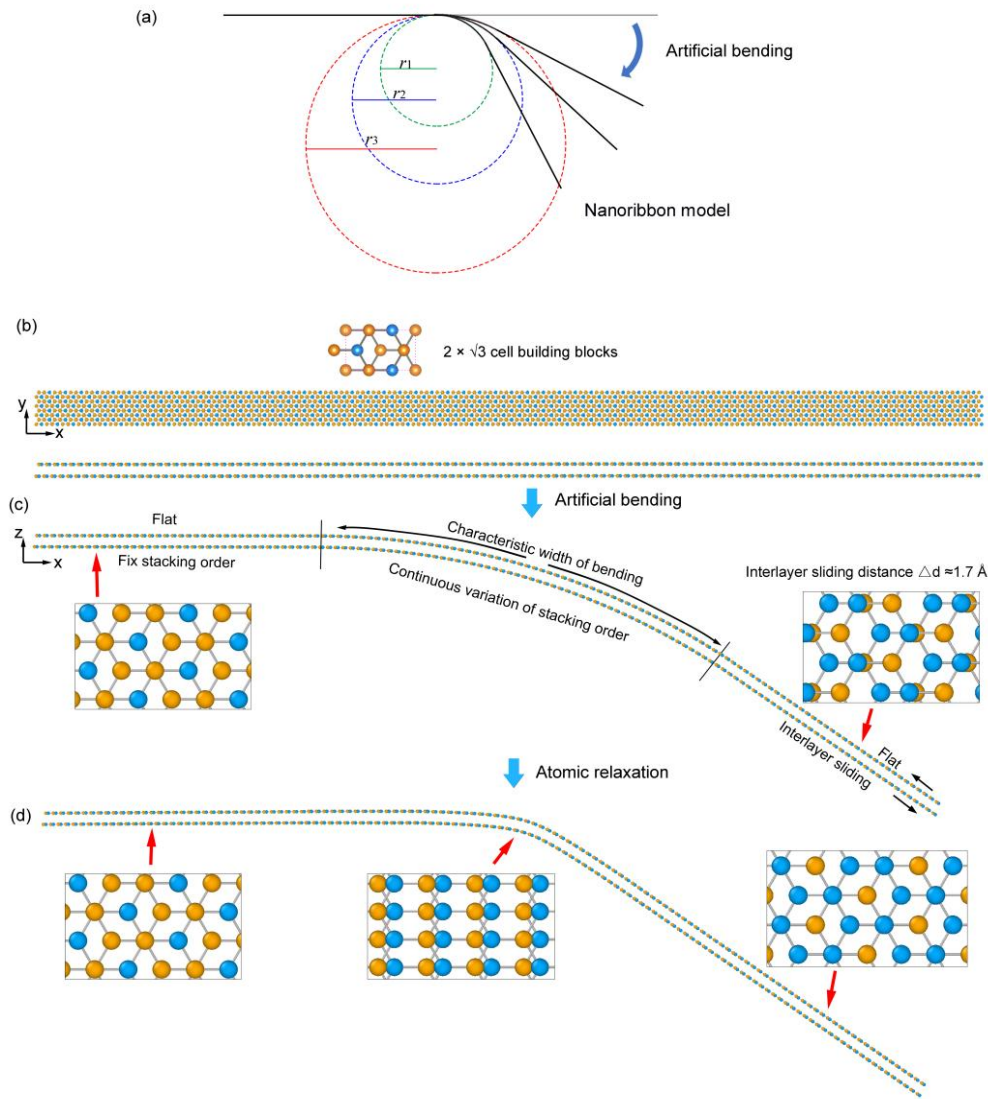

Figure S1. The bending strategy in bilyaer h-BN. (a) The sketch of artificial bending strategy. A flat sliding ferroelectricity film was artificially bend along the armchair direction with specific angle θ. The characteristic length of the bent region is 12.3 nm, and the residual regions remain flat. After bending deformation, a (b) flat bilayer sheet forms (c) an arc with interlayer sliding. (c) The DP model for the atomic relaxation leads to the optimized bilayer structures with atomically sharp kink.

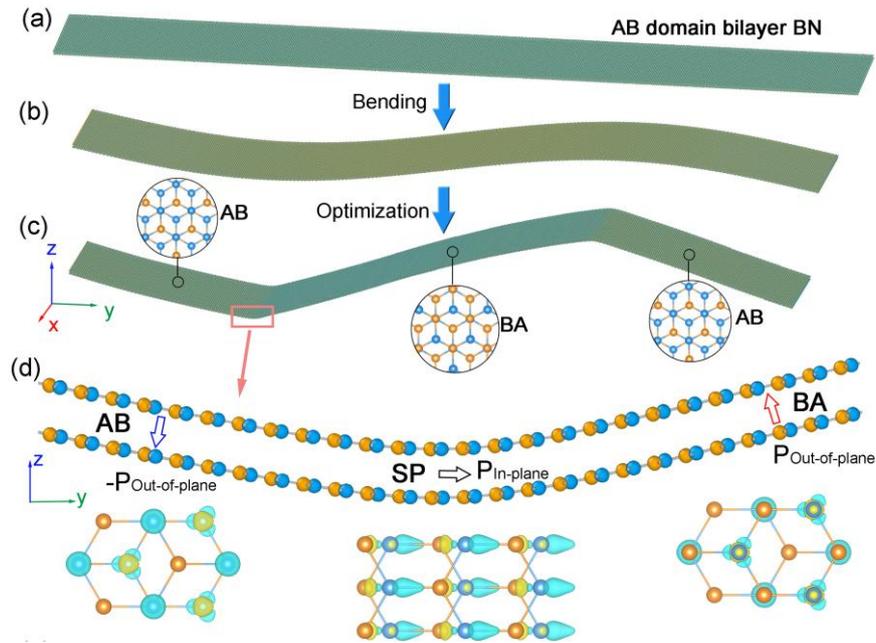

**Figure S2.** Rippling deformation in periodic sinusoidal model. (a) Bilayer flat sheet and (b) initial rippling configurations constructed by equation of $\sin(2\pi x)$. (c) Optimized structures of rippling configuration in (b) forms kinks. **Meanwhile, a interlayer sliding induced phase transition from the AB to the BA stacking occurs in the middle of the bilayer. (d) The atomic structure of cross section indicates that kink contains a** sliding ferroelectric domain wall with saddle point stacking.

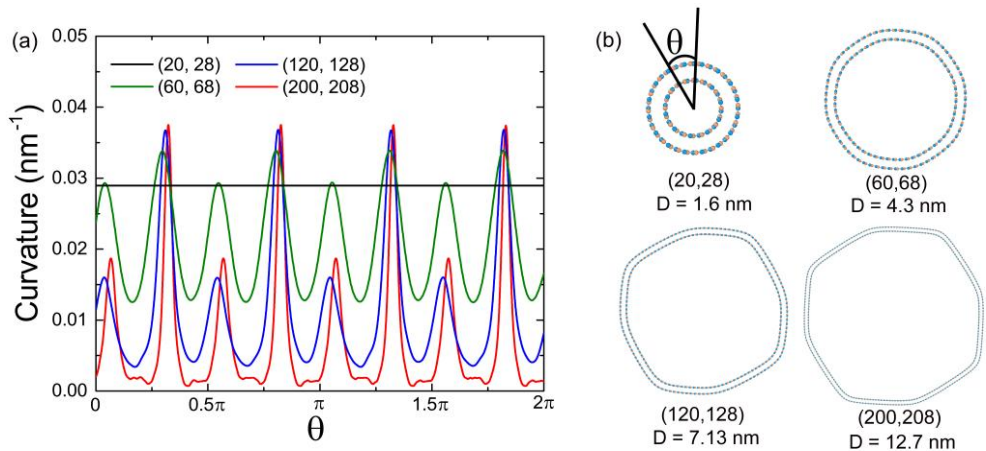

**Figure S3.** The diameter (D) dependent-geometrical morphology of double-wall nanotube h-BN. (a) The curvature distribution in the double-wall nanotube h-BN. The kinks have larger curvature in the kinked nanotube. (b) The nanotube maintains perfect circles when the diameter is less than 1.6 nm, while it transforms into kinked structure when the diameter exceeds 1.6 nm.

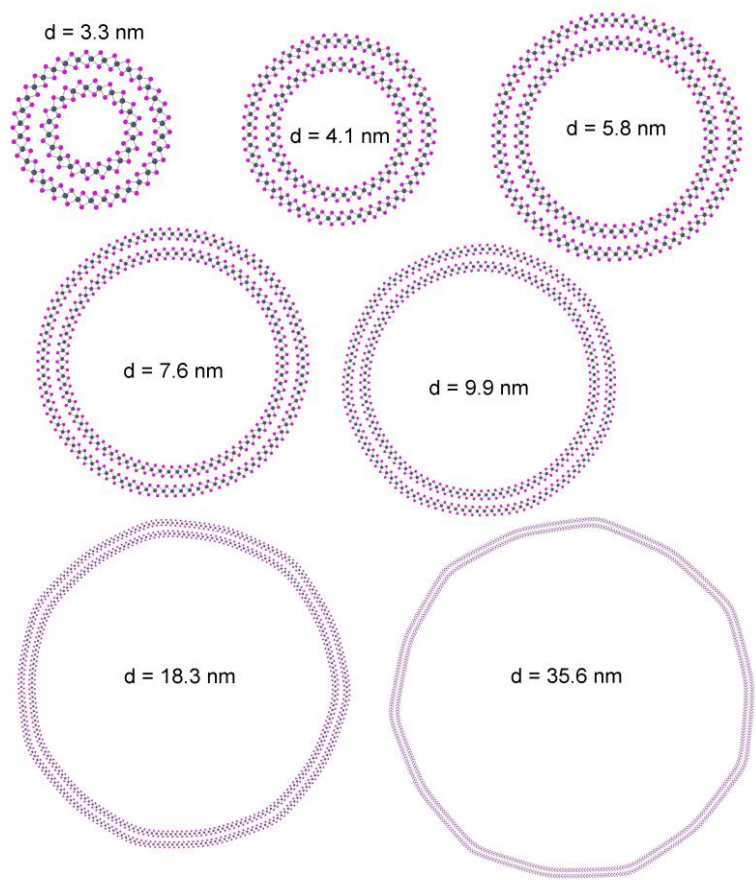

**Figure S4.** The diameter (D) of nanotube dependent geometrical morphology of double-wall nanotube 3R-MoS$_2$.

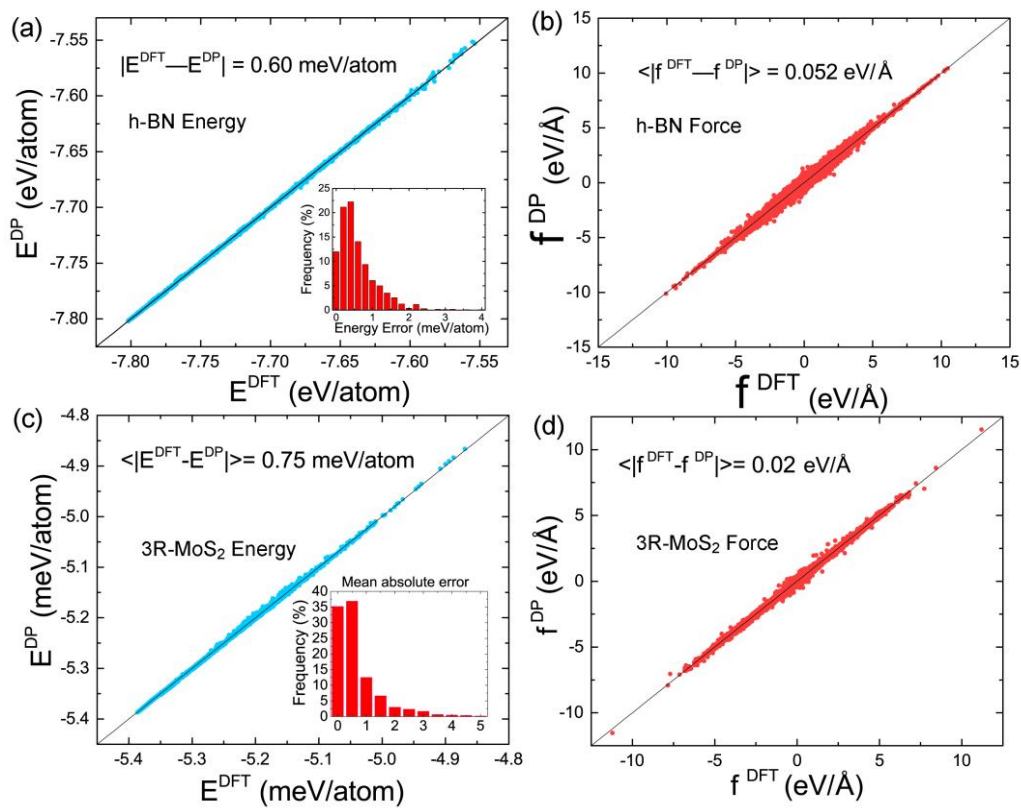

**Figure S5.** The benchmark test of the machine-learning deep potential (DP) model against the DFT calculation results. (a) (c) Comparison of energies and (b) (d) atomic force of the DP model against DFT calculations for h-BN and 3R-MoS$_2$ configurations in the training dataset. The insets show the mean absolute error of energy and between DP model and DFT results.

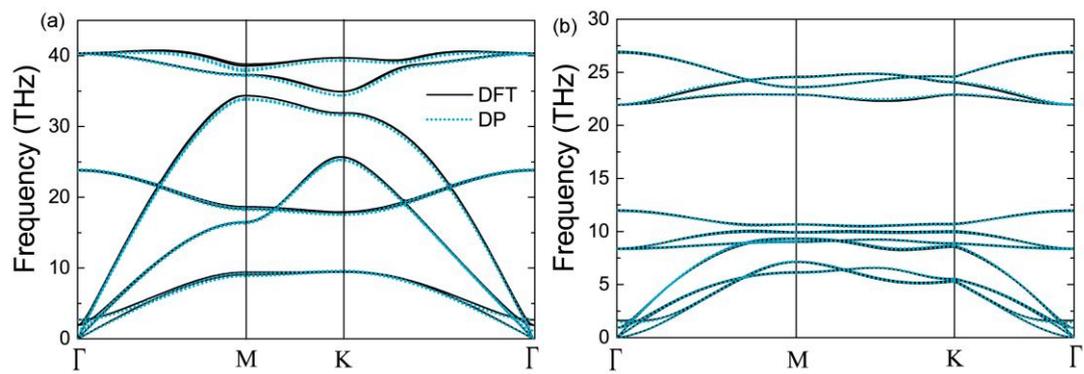

**Figure S6.** The phonon dispersion relations of the AB domain of (a) h-BN and (b) 3R-$MoS_2$ bilayers calculated by the DFT and DP model.

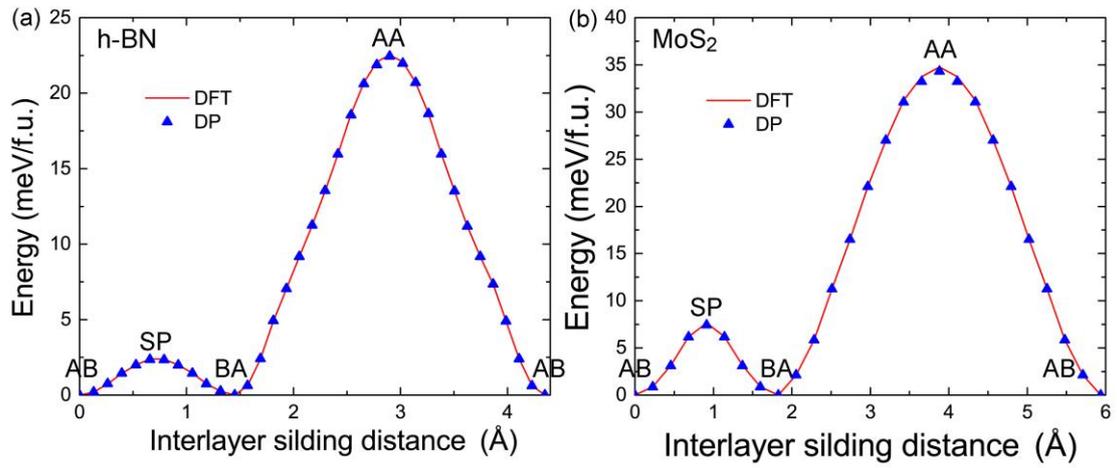

**Figure S7.** Energy changes of (a) h-BN and (b) 3R-MoS2 along the sliding pathway with minimum barrier as calculated by the DFT and DP model.

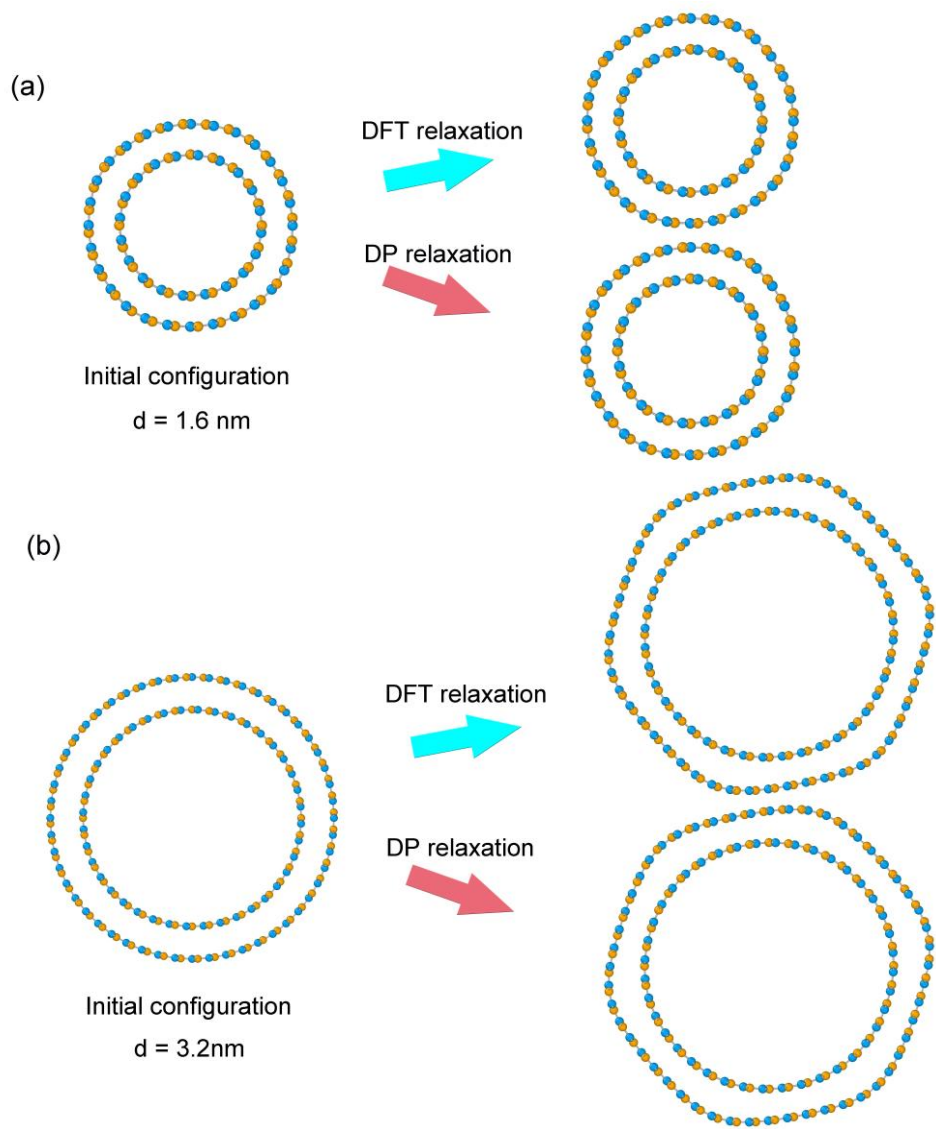

**Figure S8.** The comparison of DFT and DP relaxed atomic structures. The structural optimizations of h-BN nanotubes with diameter of (a) 1.6 nm and (b) 3.2 nm.

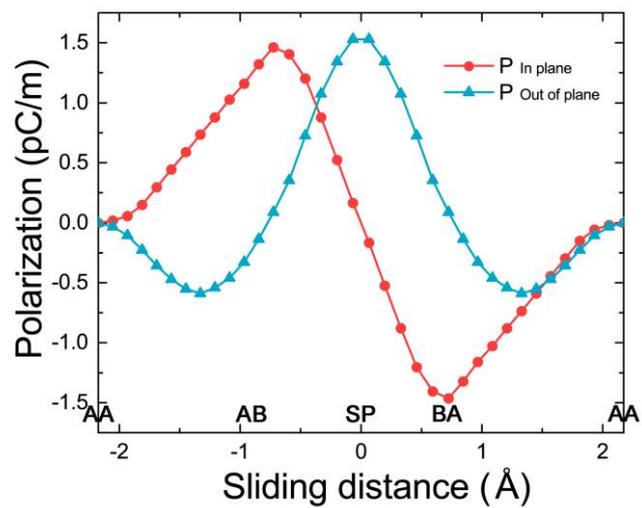

**Figure S9.** The in-plane and out-of-plane polarization profiles along the sliding pathway from AA stacking *h*-BN as calculated by the Berry phase method.

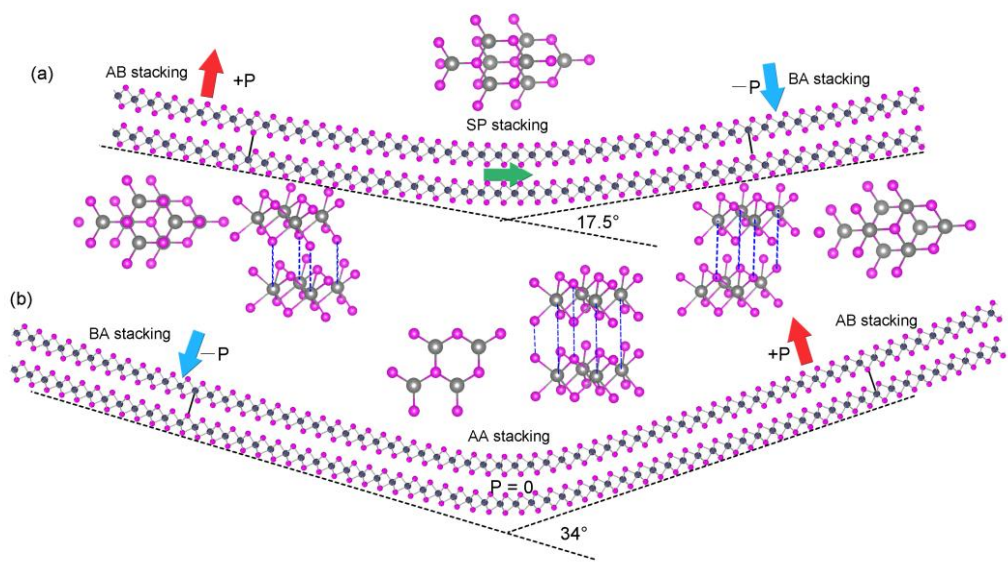

**Figure S10.** DP model optimized structures for (a) 17.5° and (d) 34° kinks in bent bilayer 3R-MoS2. These kinks form a Néel-type and Ising-like ferroelectric domain walls, as well as the h-BN case in Figure 3.

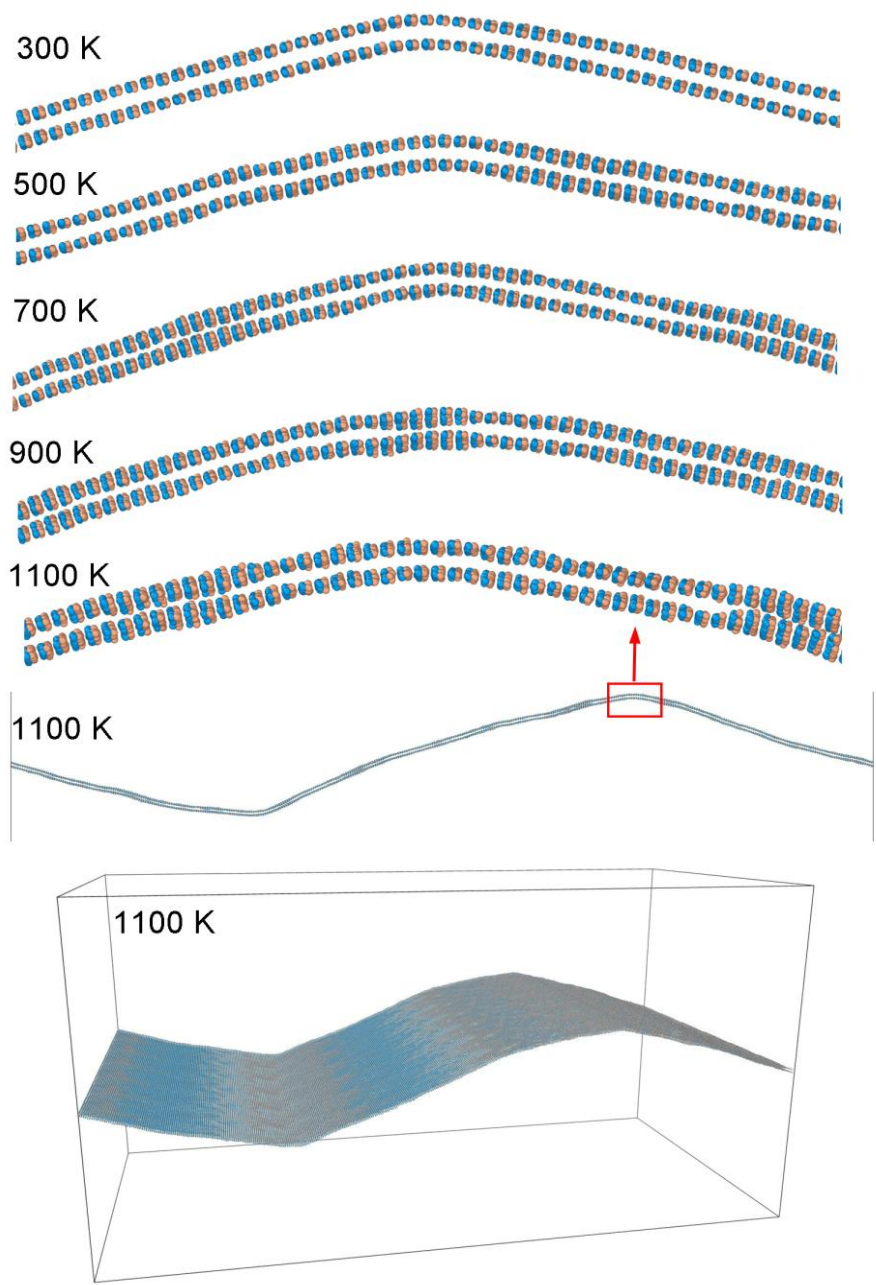

**Figure S11.** The configuration of kinks in h-BN at 300 K to 1100 K from the DP assisted molecular dynamics simulations.